%% file: main.tex
\documentclass[aps,prb,amsfonts, amssymb, amsmath, reprint, twocolumn,longbibliography]{revtex4-2}
\usepackage[english]{babel}
\usepackage{graphicx}
\usepackage{float}
\usepackage{physics}
\usepackage{epstopdf}
\usepackage{caption}
\usepackage{subcaption}
\usepackage{txfonts}
\usepackage{placeins}

\usepackage[utf8]{inputenc}
\usepackage[colorinlistoftodos, color=green!40, prependcaption]{todonotes}
\usepackage[colorlinks=true,
            citecolor=magenta,
            linkcolor=blue,
            urlcolor=blue]{hyperref}
\newcommand\varpm{\mathbin{\vcenter{\hbox{%
  \oalign{\hfil$\scriptstyle+$\hfil\cr
          \noalign{\kern-.3ex}
          $\scriptscriptstyle({-})$\cr}%
}}}}

\newcommand*{\hham}{\hat{\mathcal{H}}}

\begin{document}
\title{Curvature-Controlled Topological Magnon Phases in a Folded Kagome Lattice}

\author{Seif Alwan}
    \email{Seif.Alwan@physics.uu.se}
    \affiliation{Department of Physics and Astronomy, Uppsala University}

\author{Jonas Fransson}
    \affiliation{Department of Physics and Astronomy, Uppsala University}

\date{\today} 

\begin{abstract}
We show that geometric curvature, encoded in the folding angle between two
corner‑sharing triangles on a kagome lattice, provides a continuous tuning knob
for topological magnon phase.  Starting from an extended spin Hamiltonian with
exchange, Dzyaloshinskii–Moriya (DM) interaction, and a higher‑order bow‑tie
coupling of scalar chiralities, we derive the chirality‑mediated hopping
amplitude, which depends on the folding and spin canting of the bow-tie
triangles. At small folding and canting angles, the bow‑tie coupling surpasses
DM, establishing a curvature‑dominated regime. These results establish
curvature as an intrinsic geometric control parameter for topological magnonics
and reveal a direct analogy with chirality‑induced spin selectivity in
molecular systems, pointing to a unified mechanism for chirality driven
transport across scales. The mechanism is particularly relevant for chiral
crystals where the DM interaction is weak or forbidden by symmetry, as in
systems with a six‑fold screw axis.
\end{abstract}


\maketitle

\input{sections/introduction.tex}
\input{sections/section001.tex}  
\input{sections/section002.tex}  
\input{sections/conclusion.tex}
\input{sections/acknowledgements.tex}

\appendix*

\nocite{*}
\bibliography{refs}
\end{document}

%% file: sections/introduction.tex
\section{Introduction}\label{sec:intro}

Topological phases in magnetic materials are traditionally engineered through
relativistic spin–orbit coupling, most notably via the Dzyaloshinskii–Moriya
(DM) interaction\cite{Dzyaloshinsky_1958, Moriya_1960}, or by applying external
magnetic fields. These ingredients can generate non‑trivial band topology,
Berry curvature, and chiral edge states. In parallel, geometric curvature, long
recognized as a central concept in elasticity and optics has recently emerged
as an alternative, largely unexplored knob for controlling topological magnetic
phenomena. In geometrically frustrated magnets, where the spin texture is
exceptionally sensitive to spatial deformations, even subtle lattice folding
can dramatically alter local scalar chiralities and thereby reshape magnon
transport.
  Theoretical and experimental efforts over the past decade have established
  two primary routes to topological magnon bands in kagome lattices. The first
  relies on the DMI, which opens gaps at Dirac points and yields non-trivial
  Chern numbers, as extensively mapped in phase
  diagrams\cite{Mook2014MagnonHall, Mook2014EdgeStates} and experimentally
  verified in metal-organic framework
  compounds\cite{Chisnell2015TopologicalMagnon}. The second route employs
  long-range dipolar interactions in artificially structured magnonic crystals,
  which breaks time-reversal symmetry through magnetostatic fields and produces
  chiral edge modes\cite{Shindou2013TopologicalChiral}. Very recently, the
  interplay of DMI and pseudodipolar anisotropy has been shown to produce
  highly complex phase diagrams with high Chern
  numbers\cite{Ni2025MultipleTopological}.
  Crucially, all of these established mechanisms are rooted in either
  spin-orbit coupling or long-range electromagnetic interactions. In contrast,
  we introduce an entirely orthogonal mechanism, the geometric curvature. This
  mechanism operates independently of spin-orbit couplin and remains viable
  even when DMI is weak or symmetry-forbidden, offerin a fundamentally new
  tuning knob for topological magnonics.
However, the DM interaction itself can be severely constrained, or even
forbidden, by lattice symmetry. In chiral crystals with a high‑symmetry
stacking, such as those possessing a six‑fold screw axis, the net interlayer DM
vectors from symmetry‑equivalent bonds cancel exactly when summed over the
neighboring sites of a single magnetic ion.  This cancellation removes the
conventional route to chirality and forces us to seek alternative,
curvature‑driven mechanisms.  In the present work, we introduce one such mechanism: a
higher‑order bow‑tie interaction between scalar chiralities that is immune to
this symmetry cancellation and can operate without any spin–orbit coupling.

The kagome lattice, with its corner‑sharing triangular plaquettes as
shown in the schematic of Fig. \ref{fig:schematic_1}, provides an ideal
platform to investigate this interplay. When neighbouring triangles are folded
along a shared vertex, they form a \emph{bow‑tie} unit, and a higher‑order
chiral coupling between these bow-ties $\mathcal{K} \chi_{ijk}\chi_{ilm}$,
naturally emerges \cite{ref:Naoto_2000,ref:Taguchi2001}. The two angles that define
the geometry, the folding angle between the two triangles and the spin canting
away from the collinear state act as continuous control parameters. Tuning them
modifies both the sign and the magnitude of the resulting chirality‑mediated
hopping and drives a competition with the conventional DM exchange on the cross
bonds.

In many chiral crystals, however, the DM interaction is not only weak but may
be absent by symmetry, making a purely geometric source of chirality essential.
The bow‑tie term provides exactly such a source, its strength depends on the
curvature and canting, not on the spin–orbit coupling, thus offering a
universal design principle for chiral spin transport that can be applied even
when DM is forbidden.

In this work, we demonstrate that the geometric factor is not a small
correction but can dominate over DM interactions in a wide parameter window,
establishing a curvature‑driven topological regime. We show that the sign of
the geometric factor is intimately related to the chirality and the interaction
between the bow-ties, imprinting an interference pattern directly onto the
Chern number phase diagram. The resulting chirality‑induced edge transport
bears a striking analogy to the chirality‑induced spin selectivity
(CISS) in molecular systems\cite{ref:ciss_2012, ref:Dalum2019}, suggesting a unified picture
of geometry‑controlled topological transport across length scales.

In molecular CISS, structural chirality leads to a spin-dependent electron
transmission which gives rise to a magneto-resistance effect. Here, lattice
curvature generates handedness‑dependent magnon edge currents. Our results
therefore indicate that CISS‑like effects can emerge in solid‑state magnets
without any reference to spin–orbit coupling, broadening the scope of
chirality‑driven phenomena and offering a platform for testing the underlying
principles in a controlled lattice environment.

This study provides a minimal model that isolates the role of curvature in
topological magnonics. The insights gained are relevant for the interpretation
of recent experiments on chiral magnets such as CrNb$_3$S$_6$\cite{Togawa_2012}, where the
interplay of DM and curvature‑induced chirality determines the magnon band
structure. By establishing the geometric factor $F$ as a continuous tuning
knob, our work lays the foundation for the design of curvature‑engineered
magnonic devices, e.g., strain‑tunable topological filters or chiral magnon
waveguides.

The paper is organised as follows.  In Sec.~\ref{sec:section1} we introduce the
extended spin Hamiltonian of the kagome ferromagnet, which includes the
symmetric and asymmetric interactions extended by the chirality operators and
their interactions in the form of a bow-tie (or chirality-chirality)
interaction (see Fig. \ref{fig:schematic_1}). We determine the classical canted
ground state and then, by means of a site‑dependent Holstein–Primakoff
transformation, derive the linearised magnon Hamiltonian.  From the bow‑tie
term we extract a chirality‑mediated magnon hopping amplitude that transfers
excitation between the two outer bonds of the cluster. This coherent transfer
of chirality is the central mechanism studied in the remainder of the paper.

In Sec.~\ref{sec:section2} we analyse the classical chirality maps and the
competition with DM, then compute the topological phase diagram and edge‑state
spectra. We conclude in Sec.~\ref{sec:conclusion} with a summary and an outlook
on curvature‑engineered magnonic devices.

%% file: sections/section001.tex
\section{Method}\label{sec:section1}

\begin{figure}[t]
    \centering
    \includegraphics[height=0.7\columnwidth]{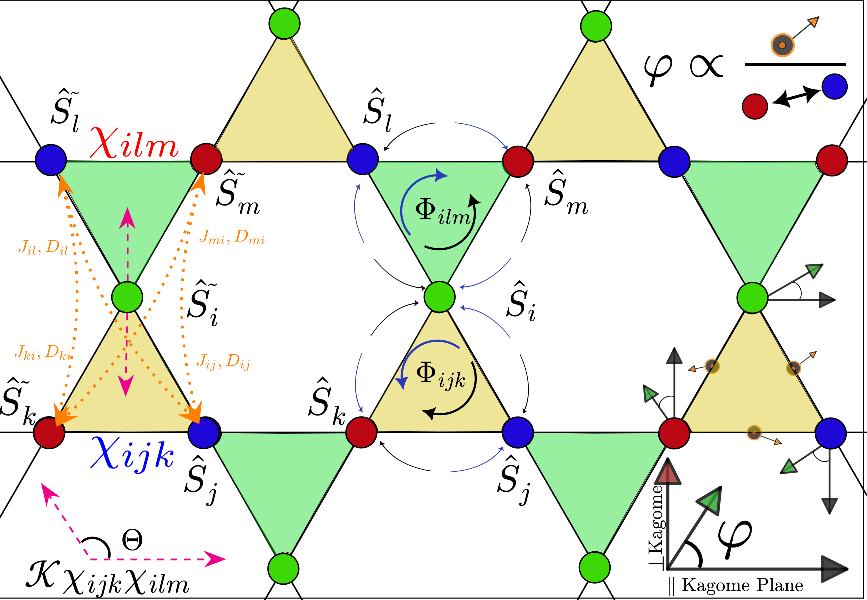}
    \captionsetup{justification=raggedright}  
    \caption{ 
      Schematic of the extended spin model on a Kagome lattice. The lattice
      consists of three sublattices represented by green, red and blue sites
      hosting $\hat{S}_{i}, \hat{S}_{j}, \hat{S}_{k}, \hat{S}_{l}$ and
      $\hat{S}_{m}$. Nearest-neighbor bonds (dashed orange arrows) are described
      by both symmetric Heisenberg exchange $J_{ij}$ and antisymmetric
      Dzyaloshinskii-Moriya (DM) interactions $D_{ij}$. The model incorporates
      scalar spin chirality terms $\hat{\chi}_{ijk}$ and $\hat{\chi}_{ilm}$
      defined on adjacent upward (yellow) and downward (green) triangular
      plaquettes, respectively. The coupling between these chiral sectors is
      represented by $\mathcal{K}\hat{\chi}_{ijk}\hat{\chi}_{ilm}$ (dashed pink
      line) with the relative canting (or twist) angle $\varphi$. The in-plane
      vector chiralities are characterized by the vortex-like vector
      orientation $\Phi_{ijk}$ and $\Phi_{ilm}$, these in turn represent the
      general handedness of the triangular plaquette. The inset (top right)
      illustrates the microscopic interplay between the symmetric Heisenberg
      exchange and the antisymmetric DMI, which governs the local
      non-collinearity. As a result of this competition, the out-of-plane spin
      canting angle $\varphi$ (defined geometrically in the bottom right inset)
      directly couples the relationship between the symmetric and antisymmetric
      exchange interactions.
    }
    \label{fig:schematic_1}
\end{figure}

We consider a two-dimensional kagome ferromagnet whose magnetic excitations
arise from coupled triangular plaquettes\cite{zhang_2020}. The system is
described by an extended spin Hamiltonian\cite{Grytsiuk_2020}
\begin{align}\label{eq:H_total}
\hat{\mathcal{H}} =&
-\frac{1}{2}\sum_{ij} J_{ij}\hat{\mathbf{S}}_{i} \cdot \hat{\mathbf{S}}_{j}
- \frac{1}{2}\sum_{\langle ij \rangle}	
	\mathbf{D}_{ij} \cdot \Bigl(\hat{\mathbf{S}}_{i} \times \hat{\mathbf{S}}_{j}\Bigr)
\nonumber
- \mu_B \mathbf{B}\cdot\sum_i \hat{\mathbf{S}}_i
\\&
	-\kappa^\text{TO}\mathbf{B}_{\text{TO}} \cdot
	\sum_{ijk}\hat{\mathbf{e}}_{ijk}
		\Bigl[ \hat{\mathbf{S}}_{i}\cdot
          		\hat{\mathbf{S}}_{j}\times \hat{\mathbf{S}}_{k}\Bigr]
              + \mathcal{K} \sum_{\langle ijk, ilm\rangle} \hat{\chi}_{ijk}\hat{\chi}_{ilm}.
\end{align}
The first term in Eq. \eqref{eq:H_total} represents the isotropic Heisenberg
interactions, with a coupling strength $J_{ij}$,
between sites $i$ and $j$, these site indices are
the vertices of the triangular plaquettes which form the Kagome lattice
structure (cf. Fig. \ref{fig:schematic_1}). 

The second term is the anisotropic Dzyaloshinskii–Moriya (DM) interaction, with
the vector coupling $\mathbf{D}_{ij}$, this expression contains antisymmetric
spin exchange. The DM interaction is followed by the Zeeman coupling, where
\(\mathbf{B}\) is the external magnetic field, and \(\mu_{B}\) is the Bohr
magneton. The sum over $\langle ij \rangle$ runs over all nearest-neighbour
bonds of the kagome lattice. Also, note that $\mathbf{B}_{\text{TO}}$ and
$\mathbf{B}$ are distinct, $\mathbf{B}$ is the uniform external field in the
Zeeman term, while $\mathbf{B}_{\text{TO}}$ is the field coupled to the
topological orbital term.

\subsubsection*{Chirality operators and the bow-tie interaction}
\label{subsec:chirality_ops}

In Eq. \eqref{eq:H_total}, the fourth term accounts for the spin chirality
\begin{align}\label{eq:chi_def}
  \hat{\chi}_{ijk} = \hat{\mathbf{S}}_i \cdot \Bigl(\hat{\mathbf{S}}_{i} \times \hat{\mathbf{S}}_{j}\Bigr), 
\end{align} 
and its coupling to the external field \( \mathbf{B}_{\text{TO}} = B_{\text{TO}}\hat{\mathbf{z}} \)
through the topological orbital susceptibility \( \kappa^\text{TO} \). The
topological orbital susceptibility $\kappa^{\text{TO}}$, represents the correlation
between the orbital magnetization operator \((\partial \hat{\mathcal{H}} /
\partial B)\) and the scalar spin chirality operator \((\partial
\hat{\mathcal{H}} / \partial \hat{\chi})\). 

The chirality operator term in Eq. \eqref{eq:H_total} represents the
topological orbital response of the electronic system to a noncoplanar spin
texture. This noncoplanar spin texture is then quantified into an effective
magnetic field, which is govenerned by the strength $\kappa^{\text{TO}}$. Note that Eq.
\eqref{eq:chi_def} tends to zero if the triple product form a coplanar
structure. Microscopically, when conduction electrons propagate through a
noncoplanar magnetic background, the scalar spin chirality acts as an emergent
magnetic flux through triangular plaquettes. 

In the present work, we are interested in the magnetic excitations in the
magnon basis. Within this effective description, \(\kappa^\text{TO}\) is
reduced to a coefficient that encodes the orbital response of the underlying
electronic system and acts as a coupling constant in the resulting magnon
theory. A microscopic derivation of this response in terms of Berry curvature
and Green's functions can be found in \cite{ref:kappaTO}. 
This emergent magnetic flux is constrained uniaxially with a unit normal vector
\(\hat{\mathbf{e}}_{ijk} = \pm \hat{\mathbf{z}}\), which points out of the
kagome plane. 
The sign (\(\pm\hat{\mathbf{z}}\)) convention (which follows the cyclic order
of the spins, as seen in Fig. \ref{fig:schematic_1}) ensures that the scalar
chirality behaves as a pseudoscalar under spatial inversion and time reversal. 

In particular, under time reversal all spins reverse sign \(\mathbf{S}_{i} \to
-\mathbf{S}_{i}\), implying \(\hat{\chi}_{ijk} \to - \hat{\chi}_{ijk}\). Therefore the
chirality \(\hat{\chi}_{ijk}\) is odd under time reversal ( $ \hat{\chi}_{ijk} \to
-\hat{\chi}_{ijk} $ ), and the corresponding term in $\hham$ {\bf explicitly breaks
time-reversal symmetry}, which is consistent with its interpretation as an
emergent magnetic flux. This flux field $\Phi_{\chi_{ijk}}$ (see Fig.
\ref{fig:schematic_1} (left)) encodes the directionality of the chiral
operator, suspended volume generated from a finite chiral contribution results
in a field centered in the triangle, inversion of this directional order
inverses the field. The scalar triple product (see Eq. \eqref{eq:chi_def}, a field
$\Phi_{\chi_{ilm}}$ is also generated from $\chi_{ilm} \neq 0$) captures the
scalar chirality and measures the signed volume spanned by three spins on a
triangular plaquette. It therefore provides a natural measure of the
noncoplanarity of the local magnetic texture.

Finally, the remaining term \(\hat{\mathcal{H}}_{\text{bow}} =
\mathcal{K}\sum_{\langle ijk, ilm\rangle} \hat{\chi}_{ijk}\hat{\chi}_{ilm}\), represents a
higher order "bow-tie" contribution to the interaction between adjacent
triangular plaquettes \( (i, j, k) \) and \( (i, l, m) \) connected through a
common vertex \( i \), modulated by \(\mathcal{K}\). This term captures the
coupling between local scalar chiralities, i.e., the interaction between
$\hat{\chi}_{ijk}$ and $\hat{\chi}_{ilm}$ as shown in Fig. \ref{fig:schematic_1}. This
expression, which is commonly referred to as the \emph{bow-tie} interaction
links pairs of adjacent triangles sharing a vertex i, thereby provides a
natural route to encode system's local curvature and noncoplanariy into the
effective magnetic interactions.  It can be thought of as an interaction
between the emergent fluxes of the two plaquettes, and it is the term that
directly transfers chirality fluctuations between the triangles. The magnitude
\(\mathcal{K}\) controls the strength of this coupling, while the folding and
canting of the triangles modulate it through a geometric factor
\(F\) which we derive below.


Having defined the chirality operators, we now proceed to bosonize the full
Hamiltonian using the Holstein-Primakoff (HP) transformation, carefully
treating the non-collinear classical ground state. However, before we can
proceed with the HP transform, we first need to define the local frame basis,
which to apply canting and folding to the system. 

\subsubsection*{Canted classical ground state}
The classical ground state of the Hamiltonian in Eq. \eqref{eq:H_total} with
ferromagnetic $J > 0$ and an out-of-plane field $\mathbf{B} =
B\hat{\mathbf{z}}$ is not simply aligned along the field due to the DM
interaction. For the symmetry-allowed DM vector $\mathbf{D}_{ij} =
D\hat{\mathbf{z}}$, the energy is minimized as the system drives towards a
non-coplanar structure. On the kagome lattice, this results in a canted
ferromagnet, spins on the three sublattices tilt uniformly away from the
$\hat{z}$-axis by a canting angle $\varphi$, but with their in-plane components
oriented at relative $2\pi/3$ angles, forming a three-sublattice chiral
structure. This configuration possesses a finite scalar spin chirality. The
angle $\varphi$ is found by minimizing the classical energy per spin, i.e.,
\begin{align}
E_{\text{cl}} =&
-\frac{1}{2}\sum_{ij}J_{ij}\hat{\mathbf{n}}_i\cdot\hat{\mathbf{n}}_j -
\frac{1}{2}\sum_{\langle ij\rangle}\mathbf{D}_{ij}\cdot(\hat{\mathbf{n}}_i\times\hat{\mathbf{n}}_j)
- \mu_B B\sum_i \hat{n}_i^z
\end{align}
 for the $ 2\pi /3$ canted structure yields , 
\begin{equation}\label{eq:canting_angle}
  \tan\varphi = \frac{D}{J + \mu_{B}B/S}.
\end{equation}
Where in the strong field limit, $\varphi \approx D/(J + \mu_{B}B/S)$. The
canting angle $\varphi$ results from the competition between the Heisenberg
exchange (which favours alignment with $\mathbf{B}$) and the DM interaction,
which favours a perpendicular components, leads to the canting angle Eq.
\eqref{eq:canting_angle}.
\subsubsection*{Site-dependent rotation}
To correctly perform linear spin-wave theory around this canted ground states,
we apply a sublattice dependent rotation $\hat{\mathcal{R}}_{p}(\varphi)$ to
each spin, aligning the new local quantization axis $(\tilde{z}_{p})$ with its
classically canted direction. For a spin on sublattice $p$, the rotation is
performed around an axis $\hat{\alpha}_{p}$ lying in the kagome plane
perpendicular to the plane spanned by $\hat{z}$ and the canted spin
direction,
\begin{subequations}
\begin{align}
  \hat{\mathbf{S}}_{i} \rightarrow \tilde{\mathbf{S}}_{i} =&
	\hat{\mathcal{R}}_{p}(\varphi)\hat{\mathcal{S}}_{i}\hat{\mathcal{R}}_{p}^{\dagger}(\varphi)
	,
\\
\hat{\mathcal{R}}_{p}(\varphi) =& \exp(-i\varphi
\hat{\alpha}_{i}\cdot \hat{\mathcal{J}}_{i}).
\end{align}
\end{subequations}
Here, $\hat{\mathcal{J}}_{i} = (\hat{J}_{i},\hat{J}_{j},\hat{J}_{k})$ is the
vector of the spin angular momentum generators satisfying
$[\hat{J}_{i}^{\mu}, \hat{J}_{j}^{\nu}] = i\delta_{ij}\epsilon_{\mu \nu
\gamma}\hat{J}_{i}^{\gamma}$. Expanding the exponential gives the series, 
\begin{equation}\label{eq:rotation_op}
  e^{i\varphi \hat{\mathcal{J}}_{\alpha}} = \mathcal{I} + i\varphi \hat{\mathcal{J}}_{\alpha} - \frac{\varphi^{2}}{2}\hat{\mathcal{J}}_{\alpha}^{2},
\end{equation}
where $\hat{\mathcal{J}}_{\alpha}$ is the generator of rotation around axis $\alpha$.

\subsubsection*{Magnon Hamiltonian in the rotated frame}
\label{subsubsec:magnon_rotated}

After the site‑dependent rotation, the classical spin at each site points along
the local $+z'$ axis.  We now introduce Holstein–Primakoff (HP) bosons for the
rotated spin operators $\tilde{\mathbf{S}}_r$, defined by the exact transformation
\begin{subequations}
\label{eq:HP_exact}
\begin{align}
\tilde S_r^{+} =& ~\left( \sqrt{2S - \hat{a}_r^\dagger\hat{a}_r}\;\right) \hat{a}_r
	,
\\
\tilde S_r^{-} =& ~\hat{a}_r^\dagger \left(\sqrt{2S - \hat{a}_r^\dagger\hat{a}_r}\right)
	,
\\
\tilde S_r^{z} =& ~S - \hat{a}_r^\dagger\hat{a}_r
	,
\end{align}
\end{subequations}
where $\hat{a}_r^\dagger$ and $\hat{a}_r$ satisfy canonical bosonic commutation
relations $[\hat{a}_r,\hat{a}_r^\dagger]=1$.  The bosonic vacuum $\ket{0}$ is the
fully ordered canted configuration, i.e.\ $\tilde S_r^{z}\ket{0}=S\ket{0}$.

For the linear spin‑wave theory used in this work, we expand the square roots
to leading order in $1/S$,
\begin{equation}
\sqrt{2S - \hat{a}_r^\dagger\hat{a}_r} \simeq \sqrt{2S}\Bigl(1 - \frac{\hat{a}_r^\dagger\hat{a}_r}{4S} + \cdots\Bigr)
\approx \sqrt{2S},
\end{equation}
which yields the familiar linearised HP approximation
\begin{subequations}
\label{eq:HP_linear}
\begin{align}
\tilde S_r^{+} \approx&
	 \sqrt{2S}\,\hat{a}_r
	 ,
\\
\tilde S_r^{-} \approx&
	\sqrt{2S}\,\hat{a}_r^\dagger
	 ,
\\
\tilde S_r^{z} \approx&
	S - \hat{a}_r^\dagger\hat{a}_r
	.
\end{align}
\end{subequations}
All subsequent expressions in this paper are derived using this linearised form.
  Higher-order corrections may quantitatively renormalize the gap but are not
  expected to alter the topological phase boundaries, which are protected by the
  sign of $F$ (see Eq. \eqref{eq:F_analytic}).
To express the spin operators in the original (global) frame, we need the
explicit matrix that relates the rotated and global coordinates.  At each site
$r$ we introduce a right‑handed orthonormal triad
${\hat{\mathbf{e}}_1^{(r)},\hat{\mathbf{e}}_2^{(r)},\hat{\mathbf{n}}_{r}}$,
where $\hat{\mathbf{n}}_r$ is the classical spin direction.  The complex
transverse vectors are $\hat{\mathbf{e}}_\pm^{(r)} = \hat{\mathbf{e}}_1^{(r)}
\pm i\hat{\mathbf{e}}_2^{(r)}$. The spin operator in the global frame is
related to the rotated spin operator as shown in Eq. \eqref{eq:rotation_op},
and the fluctuation $\delta\hat{\mathbf{S}}_r = \hat{\mathbf{S}}_r -
S\hat{\mathbf{n}}_r$ takes the compact form
\begin{equation}
  \delta\hat{\mathbf{S}}_r = \sqrt{\frac{S}{2}}\Bigl(
  \hat{\mathbf{e}}_+^{(r)} \hat{a}_r + \hat{\mathbf{e}}_-^{(r)} \hat{a}_r^\dagger \Bigr),
  \label{eq:deltaS_global}
\end{equation}
where we have used the fact that the classical part $S\hat{\mathbf{n}}_r$ is
subtracted. The factor $\sqrt{S/2}$ (rather than $\sqrt{2S}$) arises because
the transverse vectors $\hat{\mathbf{e}}_\pm^{(r)}$ have norm $\sqrt{2}$. This
normalisation ensures the correct bosonic commutation relations.

\begin{figure}[t]
    \centering
    \includegraphics[height=0.38\columnwidth]{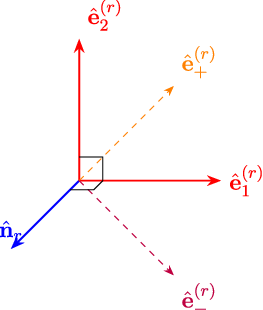}
    \captionsetup{justification=raggedright}  
    \caption{
      Local orthonormal triad at a site $r$. The classical spin
      direction $\hat{\mathbf{n}}_{r}$ (blue) defines the local $z$‑axis. The
        transverse unit vectors
        $\hat{\mathbf{e}}_1^{(r)},\hat{\mathbf{e}}_2^{(r)}$ (red) span the
        perpendicular plane. The complex combinations
        $\hat{\mathbf{e}}_\pm^{(r)} = \hat{\mathbf{e}}_1^{(r)}\pm
      i\hat{\mathbf{e}}_2^{(r)}$ (orange and purple dashed arrows) are used to
    express the Holstein–Primakoff fluctuation. Orthogonality is indicated by
  the small right‑angle marks.
    }
    \label{fig:local_spin_frame}
\end{figure}

A convenient choice for the local triad that respects the $2\pi/3$ kagome
geometry is
\begin{subequations}
\label{eq:local_triad}
\begin{align}
\hat{\mathbf{e}}^{(r)}_1 =&
	\begin{pmatrix}
		\cos\varphi\cos(\theta_r+\varphi) \\
		\cos\varphi\sin(\theta_r+\varphi) \\
		-\sin\varphi
	\end{pmatrix}
	,
\\
\hat{\mathbf{e}}^{(r)}_2 =&
	\begin{pmatrix}
		-\sin(\theta_r+\varphi) \\
		\cos(\theta_r+\varphi) \\
		0
	\end{pmatrix}
	,
\\
\hat{\mathbf{n}}^{(r)} &= \hat{\bf e}^{(r)}_1\times\hat{\bf e}^{(r)}_2
	,
\end{align}
\end{subequations}
which spans the plane transverse to the classical spin
direction.  For the first triangle the intrinsic azimuths are $\theta_i = 0$,
$\theta_j = 0$, $\theta_k = 2\pi/3$.  For the second triangle, the folding
shifts these angles to $\theta_l = \pi$ and $\theta_m = \pi+2\pi/3$ at
$\Theta=0$, the spherical‑arc folding (described in the Supplementary Material)
continuously deforms them as $\Theta$ increases.
Figure~\ref{fig:local_spin_frame} illustrates the local triad.

We now replace the spin operators in the full Hamiltonian~\eqref{eq:H_total} by
$S\hat{\mathbf{n}}^{(r)} + \delta\hat{\mathbf{S}}_r$ with the fluctuation given by
\eqref{eq:deltaS_global} and retain only terms quadratic in the bosons, that is, the linear
spin‑wave approximation. The Heisenberg, DM, and Zeeman parts are standard,
and after the site‑dependent rotation they take the compact forms
\begin{align}
\hat{\mathcal{H}} =&
-\frac{S}{2}\sum_{ij}
	\Bigl(
		J_{ij}+iD_{ij}\Bigr)
		\Bigl(\hat{a}_i^\dagger\hat{a}_j+\text{h.c.}\Bigr)
	-\mu_B S B\sum_i \hat{a}_i^\dagger\hat{a}_i
\nonumber\\&
  - B_{\text{TO}}\kappa^{\text{TO}}\!\!\sum_{\mathrm{CW,CCW}}\!\hat{\chi}_{ijk}
   + \mathcal{K}\!\sum_{\langle ijk,ilm\rangle}\!\hat{\chi}_{ijk}\hat{\chi}_{ilm}.
\label{eq:H_magnon_rot}
\end{align}
The DM coupling $D_{ij}^{z}$ is the out‑of‑plane component of the rotated DM
vector since the in‑plane components do not contribute at quadratic order.  The
chirality operators $\hat{\chi}_{ijk}$ appearing in the last two terms are now
understood to be the \emph{rotated} chirality operators, whose explicit
linearised form after rotation is
\begin{equation}
  \hat{\chi}_{ijk} = iS^{2}\Big[
  (\hat{a}_i^\dagger\hat{a}_j - \hat{a}_j^\dagger\hat{a}_i)
+ (\hat{a}_j^\dagger\hat{a}_k - \hat{a}_k^\dagger\hat{a}_j)
+ (\hat{a}_k^\dagger\hat{a}_i - \hat{a}_i^\dagger\hat{a}_k)
  \Big]
  \cos2\varphi.
  \label{eq:chi_rotated_final}
\end{equation}
The factor $\cos2\varphi$ originates from the projection of the rotated spin
components.

Equation~\eqref{eq:H_magnon_rot} together with \eqref{eq:chi_rotated_final}
constitutes the single‑magnon Hamiltonian used throughout the paper. The
four‑spin bow‑tie term will be further expanded in the next subsection to
extract the geometric factor $F$ that encodes the folding angle dependence.


\subsubsection*{Geometric $\Gamma$ coefficients}

We can now proceed to define geometric coefficients,
\begin{subequations}
\begin{align}
  \Gamma_{ijk}(i, \pm) =&
  	 \hat{\mathbf{n}}_j \cdot (\hat{\mathbf{n}}_k \times \hat{\mathbf{e}}_\pm^{i})
\\
  \Gamma_{ijk}(j, \pm) =&
  	 \hat{\mathbf{n}}_k \cdot (\hat{\mathbf{n}}_i \times \hat{\mathbf{e}}_\pm^{j})
\\
  \Gamma_{ijk}(k, \pm) =&
  	 \hat{\mathbf{n}}_i \cdot (\hat{\mathbf{n}}_j \times \hat{\mathbf{e}}_\pm^{k})
\end{align}
\end{subequations}
These $\Gamma$ coefficients are purely geometric, they depend only on the
classical directions and the choice of transverse basis. They encode the
coupling between the classical spin orientations and the quantum fluctuations.
The complex conjugate relation is $\bar{\Gamma}_{ijk}(p, +) = \Gamma_{ijk}(p,
-)$, because $\hat{\mathbf{e}}_+ $ and $\hat{\mathbf{e}}_- $ are complex
conjugates while the other vectors are real.

The full chirality operator is $\hat{\chi}_{ijk} = \hat{\chi}_{ijk}^{(0)} +
\hat{\chi}_{ijk}^{(1)}$, where the classical part $\hat{\chi}_{ijk}^{(0)} =
S^{3}\hat{\mathbf{n}}_{i}\cdot (\hat{\mathbf{n}}_{j} \times
\hat{\mathbf{n}}_{k})$ is purely geometric. The classical product map discussed
in ~\ref{sec:section2} uses the classical expression $\hat{\chi}^{0}_{ijk}$.
The quantum fluctuation, $\hat{\chi}_{ijk}^{(1)}$, can be expressed through the
chirality operator in terms of geometric coefficients, 
\begin{align}\label{eq:chi_gamma}
  \hat {\chi}^{(1)}_{ijk} = S^{2}\sqrt{\frac{S}{2}}\sum_{p\in \{ i,j,k \}}
  \left[\Gamma_{ijk}(p,+)\hat{a}_{p} +
  \Gamma_{ijk}(p,-)\hat{a}_{p}^{\dagger}\right] + h.c.
\end{align}
Here, $\hat{\chi}^{(1)}_{ijk}$ is linear in the quantum fluctuations, and
contains single-boson creation and annihilation operators weighted by geometric
coefficients $\Gamma_{ijk}(p, \pm)$. The $\Gamma$ coefficients formalism shown
here reproduces the $\cos2\varphi$ factor in the unfolded geometry.

The bow-tie interaction, Eq. \eqref{eq:H_magnon_rot}, couples two triangles
through the term $\mathcal{K}\hat{\chi}_{ijk}\hat{\chi}_{ilm}$. Because each
$\hat{\chi}$ is linear in the bosons, their product is quadratic.
Using Eq. \eqref{eq:chi_gamma} for both triangles we obtain,
\begin{align}
\hat{\chi}_{ijk}\hat{\chi}_{ilm} =&
	 -2S^{5}\sum_{p \in \{i,j,k \}} ~ \sum_{q \in \{i,l,m \}}
	\Bigl(\Gamma_{ijk}(p,+)\Gamma_{ilm}(q,+)\hat{a}_{p}\hat{a}_q
\nonumber\\&
	+ \Gamma_{ijk}(p,+)\Gamma_{ilm}(q,-)\hat{a}_{p}\hat{a}_q^{\dagger}
	+ \Gamma_{ijk}(p,-)\Gamma_{ilm}(q,+)\hat{a}_{p}^{\dagger}\hat{a}_q 
\nonumber\\&
	+ \Gamma_{ijk}(p,-)\Gamma_{ilm}(q,-)\hat{a}_{p}^{\dagger}\hat{a}_q^{\dagger}
	\Bigr) + h.c.
\end{align}
Here, the first and last terms account for the \emph{anomalous} two magnon
annihilation and creation, respectively, which will be omitted in the following
since they create/destroy two magnons simultaneously and do not contribute to
the single-magnon Hamiltonian at leading order in $1/S$. 
The focus instead lies on the second and third terms, which describe magnon
exchange between the first triangle $(i, j, k)$ (the upward blue triangle and
second triangle $(i, l, m)$ (folded red triangle) (see Fig.
\ref{fig:schematic_1}).
%
To obtain the effective single-particle Hamiltonian that
governs magnon propagation, we normal-order the product with respect to the
unperturbed ground state $\ket{0}$, where $\ket{0}$ is the magnon vacuum of the
canted ground state. Schematically, $\hat{a}_{p}\hat{a}_{q}^{\dagger} =
\hat{a}_{q}^{\dagger}\hat{a}_{p} + \delta_{pq}$, so that the normal-ordered
hopping terms are equal to, e.g., $\hat{a}_p^{\dagger}\hat{a}_q$ up to a constant. The latter merely renormalise the
classical ground-state energy and do not affect the single-magnon spectrum.
By, furthermore, restricting to terms
that involve the outer sites $j, k$ and $l, m$ (because the central vertex $i$
contributes only to diagonal on-site energies), we arrive at

\begin{align}
\label{eq:geo_chichi}
    \hat{\chi}_{ijk}\hat{\chi}_{ilm} =&
    	 -2S^{5}\sum_{p \in \{j,k \}}\sum_{q \in \{l,m \}}
	    \Gamma_{ijk}(p,-)\Gamma_{ilm}(q,+)\hat{a}_{p}^{\dagger}\hat{a}_{q} + h.c.
\end{align}

The product of $\Gamma$ coefficients shown in Eq. \eqref{eq:geo_chichi} is a
complex number whose magnitude and phase depend on $\Theta$ and $\varphi$, and
can be evaluated analytically for the specific spherical-arc folding geometry
(see Supplementary Materials for further detailed derivation). The explicit
evaluation of the $\Gamma$ coefficients using the classical spin directions and
the local triads of Eqs. \eqref{eq:local_triad} yields the compact expression,

\begin{align}
\label{eq:F_analytic}
  F(\Theta,\varphi) =&
  	-\frac{3}{16}S^{5}
	\Bigl(
		2\cos\varphi\cos\Theta - \sin\Theta\sin\varphi
	\Bigr)
	\cos^3\varphi
	.
\end{align}
Throughout the text we will refer $F(\Theta,\varphi)$ of equation
\eqref{eq:F_analytic} simply as $F$ for simplicity. The magnitude \(|F|\)
controls the strength of the chiral‑mediated hopping, while its sign determines
the effective phase of the hopping amplitudes. Consequently, the interference
between the two triangles is constructive when \(|F|\) is large and destructive
when \(F\) changes sign or vanishes.  The zeros of \(F\) occur along the curve
$\cot\Theta = \frac12\tan\varphi$ and $F=0,$ which corresponds to
configurations where the two triangles are exactly orthogonal in terms of their
chiral coupling. Maximal constructive interference happens when \(|F|\) reaches
a local maximum, differentiating \(F\) with respect to $\Theta$ at fixed
$\varphi$ yields the condition $\tan\Theta = -\frac12\tan\varphi$. These
relations provide a direct map between the geometric parameters and the
resulting interference pattern, and they serve as analytic guides for locating
topological phase transitions.
  The idealized spherical-arc folding geometry resulting from Eq.
  \eqref{eq:F_analytic}, represents the clean limit; in real materials,
  disordered strain gradient may broaden the gap but will not destroy the
  topological protection as long as the gap remains finite.

The folding angle \( \Theta \) affects inter site distances within the bow-tie
unit, leading to curvature dependent effective couplings:
\begin{align}\label{eq:distance_law}
J_{ij} = J_0\, e^{-(d_{ij}(\Theta)-1)/\xi_J}, \qquad
D_{ij} = D_0\, e^{-(d_{ij}(\Theta)-1)/\xi_D},
\end{align}
where \( d_{ij}(\Theta) \) is the separation between spins \( i \) and \( j \),
and \( \xi_J,\xi_D \) are decay lengths for the Heisenberg and DM exchanges,
respectively. Here $J_0$ and $D_0$ are the Heisenberg exchange and DM strength
respectively, for a bond length $R = 1$ (the reference intra-triangle
distance).
This provides a parametrized form of the exchange interactions. The combined
effect of curvature and canting modifies both the magnitude of the effective
couplings and the complex phase of the magnon hopping amplitudes, giving rise
to an emergent gauge flux associated with the scalar chirality field. In the
absence of folding and canting, the system remains coplanar with \( \chi_{ijk}
= 0 \) and no topological phase emerges. Finite folding, \( \Theta \),  and
canting, \( \varphi \), generate nonzero solid angles, producing finite scalar
chirality. The chirality–chirality coupling in Eq.~\eqref{eq:geo_chichi}
thereby introduces an effective curvature mediated topological stiffness that
stabilizes chiral magnonic modes.

The bow-tie term contributes cross-triangle hoppings with bond vectors \(
\delta_{\alpha, \beta }\). In the Bloch Hamiltonian these appear as 
\begin{equation}
  [ \hham_{bow}(\mathbf{k}) ]_{\alpha \beta} = \eta_{\alpha \beta} \frac{F(\Theta, \phi)}{2} e^{i\mathbf{k}\cdot \delta_{\alpha \beta}}.
\end{equation}
Here, \(\eta_{\alpha \beta}\) is the sign factor of the cross-triangle
hopping: $ \eta_{jl} = \eta_{km} = + 1, ~ \eta_{jm} = \eta_{kl} = -1$,  for the
bonds connecting sublattice $\alpha$ to $\beta$ with displacement $\delta$, and
$\eta_{\alpha \beta} = 0$ otherwise.
The velocity operator is the $\mathbf{k}$-derivative:
\begin{equation}\label{eq:velocity_op}
  \partial_{k_\mu} H_{\text{bow}}(\mathbf{k}) = i\sum_{\alpha \beta} \frac{F(\Theta,\varphi)}{2} \delta^{\mu}_{\alpha \beta} \eta_{\alpha\beta}(\boldsymbol{\delta})\, e^{i\mathbf{k}\cdot\boldsymbol{\delta}},
\end{equation}
where the sum should run over the four cross-triangle bonds explicitly. The
index $\mu$ labels the two independent directions in the two-dimensional
momentum space $\mathbf{k}= (k_{x}, k_{y})$, and $\partial_{k_{\mu}} =
(\partial/\partial_{k_{x}}, \partial/\partial_{k_{y}})$.

We can now examine how the bow-tie interaction affects the band topology. The
Berry curvature of the $n$-th magnon band is obtained from the Bloch
Hamiltonian $\hham(\mathbf{k})$ using the standard formula for a
number-conserving bosonic system, 
\begin{align}
\Omega_n(\mathbf{k}) = -2\,\mathrm{Im} \sum_{m \neq n}
\frac{
\langle u_{n\mathbf{k}} | \partial_{k_x} \mathcal{H}(\mathbf{k}) | u_{m\mathbf{k}} \rangle
\langle u_{m\mathbf{k}} | \partial_{k_y} \mathcal{H}(\mathbf{k}) | u_{n\mathbf{k}} \rangle
}{
\left( \varepsilon_{n\mathbf{k}} - \varepsilon_{m\mathbf{k}} \right)^2
}.
\label{eq:berry_standard}
\end{align}
Where $\varepsilon_{n\mathbf{k}}$ and $\ket{u_{n\mathbf{k}}}$ are eigenvalues
and eigenvectors of the $3 \times 3$ matrix $\hham (\mathbf{k}) = \hham_{0}
(\mathbf{k}) + \hham_{\text{bow}} (\mathbf{k})$. Here all other interactions of
Eq.~\eqref{eq:H_total} are contained in $\hham_{0}$. Although this model
simplifies our original Hamiltonian, Eq.~\eqref{eq:H_total}, it does gives us
an intuitive picture of the contribution of the bow-tie interaction on the
topology of the system. The velocity operators
$\partial_{k_{\mu}}\hham$ are Hermitian matrices that describe how the
single-particle states change with momentum. Equation
~\eqref{eq:berry_standard} involves a product of off-diagonal matrix elements
of these velocity operators, the imaginary part selects the antisymmetric
(Hall-type) response. The diagonal matrix element $\langle
u_n|\partial_{k_\mu}\mathcal{H}|u_n\rangle$ is purely real (it equals
$\partial_{k_\mu}\varepsilon_n$), hence it drops out of the imaginary part,
leaving only inter‑band transitions. Although the energy differences in the
denominator also depend on $F$, this dependence is subleading when $F$ is small
compared to the gap opened by the other interactions. Thus to leading order the
bow‑tie contribution to the Berry curvature scales as $|F|^2$.
Substituting $\partial_{k_{\mu}}\hham_{\text{bow}}$ into the standard Berry curvature
formula ~\eqref{eq:berry_standard}, the numerator becomes proportional to
$|F|^{2}$. 

Hence the bow-tie induced Berry curvature scales as,  
\begin{equation}
\Omega_n^{\text{bow}}(\mathbf{k}) \propto |F(\Theta,\varphi)|^2 \times \text{(band‑structure terms)}.
\end{equation}
Here, the proportionality constant includes powers of \(S\) and the coupling
constants \(\kappa_{\text{TO}}\) and \(B_{\text{TO}}\), but the essential
geometric dependence is carried by $|F|^{2}$.  

The sign of $F$ influences the phase of the velocity matrix elements, when $F$
changes sign, each $\bra{u_{m}}\partial_{k_{\mu}}\hham_{\text{bow}}\ket{u_{n}}$
acquires a factor of $e^{i\pi}$, which can flip the sign of the imaginary part
of the product in Eq.~\eqref{eq:berry_standard}. Therefore, although the
magnitude of $\Omega_{n}^{\text{bow}}$ is determined by $|F|^{2}$, the sign of the
Berry curvature, and the resulting topological characters of the band, can be
reversed when $F$ crosses zero. 

This observation directly links the geometric angles $\Theta$ and $\varphi$ to
topological phase transitions. In the simplified model, $\hham(\mathbf{k}) =
\hham_{0}(\mathbf{k}) +\hham_{\text{bow}}(\mathbf{k})$ which we discussed above, the
bow‑tie interaction couples the two antisymmetric bond modes with an amplitude
$F$. This simplifies to,
\begin{equation}\label{eq:2-mode}
\hham = \varepsilon(\hat{b}_{1}^{\dagger}\hat{b}_{1} +
\hat{b}_{2}^{\dagger}\hat{b}_{2}) + 
F(\Theta,
\varphi)\hat{b}_{j,k}^{\dagger}\hat{b}_{l,m} + F^{*}(\Theta,
\varphi)\hat{b}_{l,m}^{\dagger}\hat{b}_{j,k}, 
\end{equation}
where $\hat{b}^{\dagger}_{j,k} = (1/\sqrt{2})(\hat{a}_{j} - \hat{a}_{k})$ and $\hat{b}_{l,m}
= (1/\sqrt{2})(\hat{a}_{l} - \hat{a}_{m})$, these are the antisymmetric
(bond-odd) operators which create and annihilate a magnon on the bonds (j,k)
and (l,m), respectively . These operators arise from the factorization of the
bow-tie operator term $\hat{\chi}_{ijk}\hat{\chi}_{ilm} \propto
S^{4}(\hat{a}_{j}^{\dagger} - \hat{a}_{k}^{\dagger})(\hat{a}_{l} - \hat{a}_{m})
+ h.c.$.
The bow-tie interaction in Eq. \eqref{eq:2-mode} is reduced to a hopping term,
$\hham_{\text{bow}} \to F(\Theta, \varphi)\hat{b}_{1}^{\dagger}\hat{b}_{2} + h.c.$. It
then becomes clear that when $F = 0$, the two modes are degenerate at energy
$\varepsilon$, any finite $F$ hybridises them into symmetric and antisymmetric
combinations split by $2|F|$. The sign of $F$ determines which combination has
the higher energy, so a sign change of $F$ therefore inverts the bands. Further
interactions in the full model of Eq.~\eqref{eq:H_total}, modify the diagonal
elements and can lift the degeneracy at $F = 0$.

Nevertheless, the sign of $F$ controls the parity of the hybridised states, as
$F$ changes sign, the symmetric and antisymmetric combinations of the two bond
modes invert, and this band inversion drives a topological transition. The
precise point where the gap closes and the Chern number changes is therefore
shifted by additional interactions from the line $\cot\Theta = \frac12
\tan\varphi$. However, the global structure of the phase diagram is still
governed by the sign change of $F$.

The topological character of each band is quantified by the total Chern number
\begin{equation} 
C_n = \frac{1}{2\pi}\int_{\mathrm{BZ}} \Omega_n(\mathbf{k})\,d^2k,
\end{equation}\label{eq:chern}
where $\Omega_n(\mathbf{k})$ is given by Eq.~\eqref{eq:berry_standard}. Because
$\Omega_n^{\mathrm{\text{bow}}}(\mathbf{k}) \propto |F|^2$, the bow‑tie‑induced Berry
curvature is maximal when $|F|$ is large and vanishes when $F=0$. At the same
time, the sign of $F$ influences the phase of the velocity matrix elements and
can reverse the sign of the Berry curvature, enabling changes in the Chern
numbers. As we cross the critical region where $F$ changes sign, the band gap
closes and reopens, and the Chern numbers $C_n$ exhibit integer jumps,
signaling topological phase transitions driven primarily by the geometric
parameters $\Theta$ and $\varphi$. In the following section \ref{sec:section2},
we will show that the interference conditions derived from $F$ are directly
reflected in the numerically computed Chern numbers, confirming that the
geometric factor controls the topological phase diagram of the kagome
ferromagnet.

%% file: sections/section002.tex
\section{Results}\label{sec:section2}

We first examine how the classical scalar chiralities of the two triangles vary
with the folding angle $\Theta$ and the canting angle $\varphi$.  The interplay
between curvature, spin canting, and chiral correlations provides a direct
geometric encoding of the quantum interference that will later be shown to
govern the topological phase diagram.

All numerical results in this section use the common parameters $S = 1$ and
$\mathcal{K} = 1$. The classical chirality maps and the analysis of the
competition between the geometric coupling $F$ and the DM interaction
(Figs.~\ref{fig:sum_prod}, \ref{fig:F_over_D}) were obtained with $D_0 =
0.3\,J_0$, $\xi_D = 0.5$, and $\xi_J = 0.6$ (taking $J_0 = 1$). The topological
phase diagram and ribbon spectra (Figs.~\ref{fig:chern_phaseDiagram} and
\ref{fig:ribbon_combined}) were computed using the parameter set $J = 0.8$, $D
= 0.6$, $B = 0.2$, $B_{\text{TO}} = 0.2$, and $\kappa^{\text{TO}} = 0.6$, which
ensures a well‑resolved topological gap.

\subsubsection*{Classical chirality landscape and competition with DM}

\begin{figure}[t]
    \centering
    \begin{subfigure}[t]{0.4\textwidth}
        \centering
        \includegraphics[width=\textwidth]{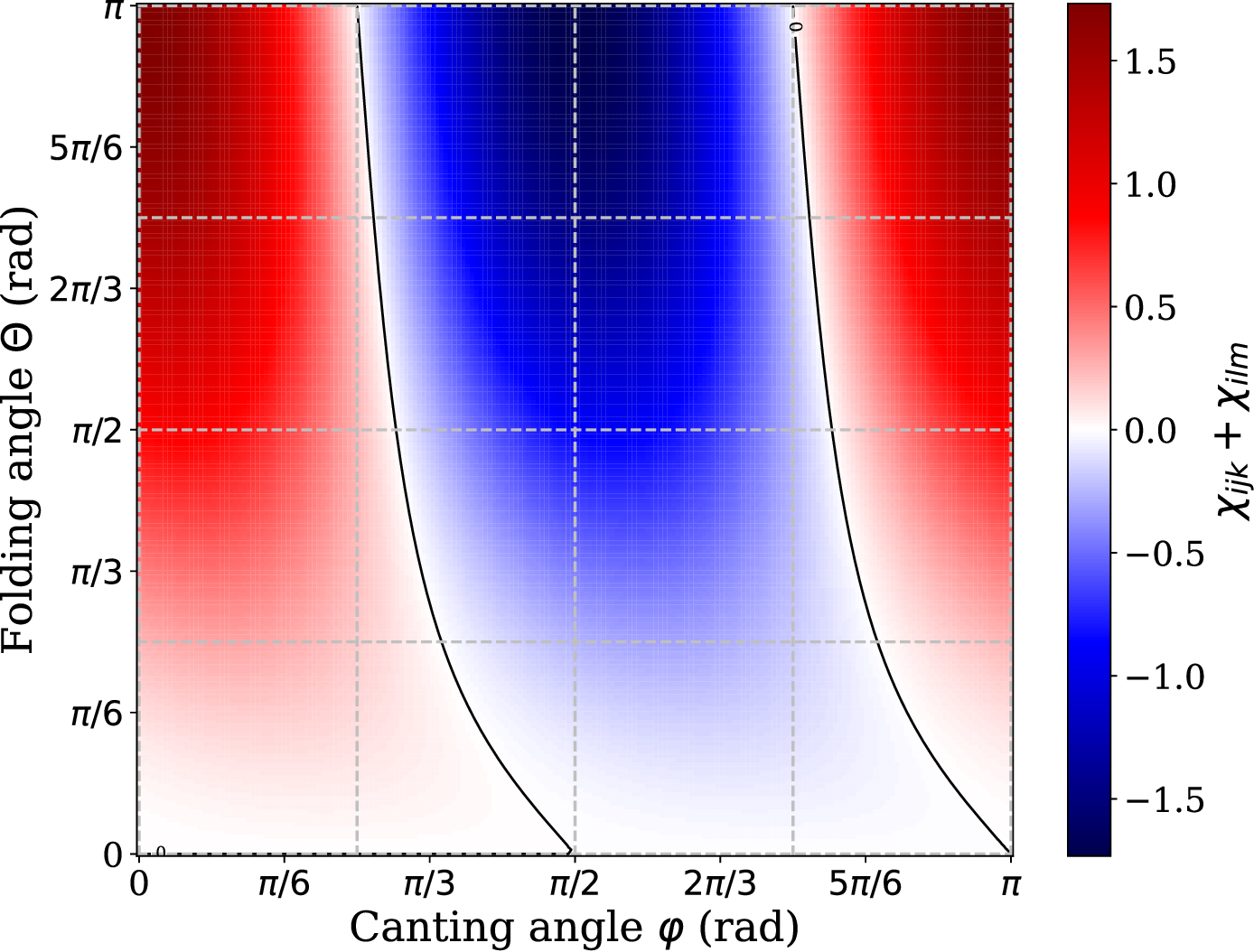}
        \caption{Sum $\chi_{ijk}+\chi_{ilm}$}
        \label{fig:sum_panel}
    \end{subfigure}
    \hfill
    \begin{subfigure}[t]{0.4\textwidth}
        \centering
        \includegraphics[width=\textwidth]{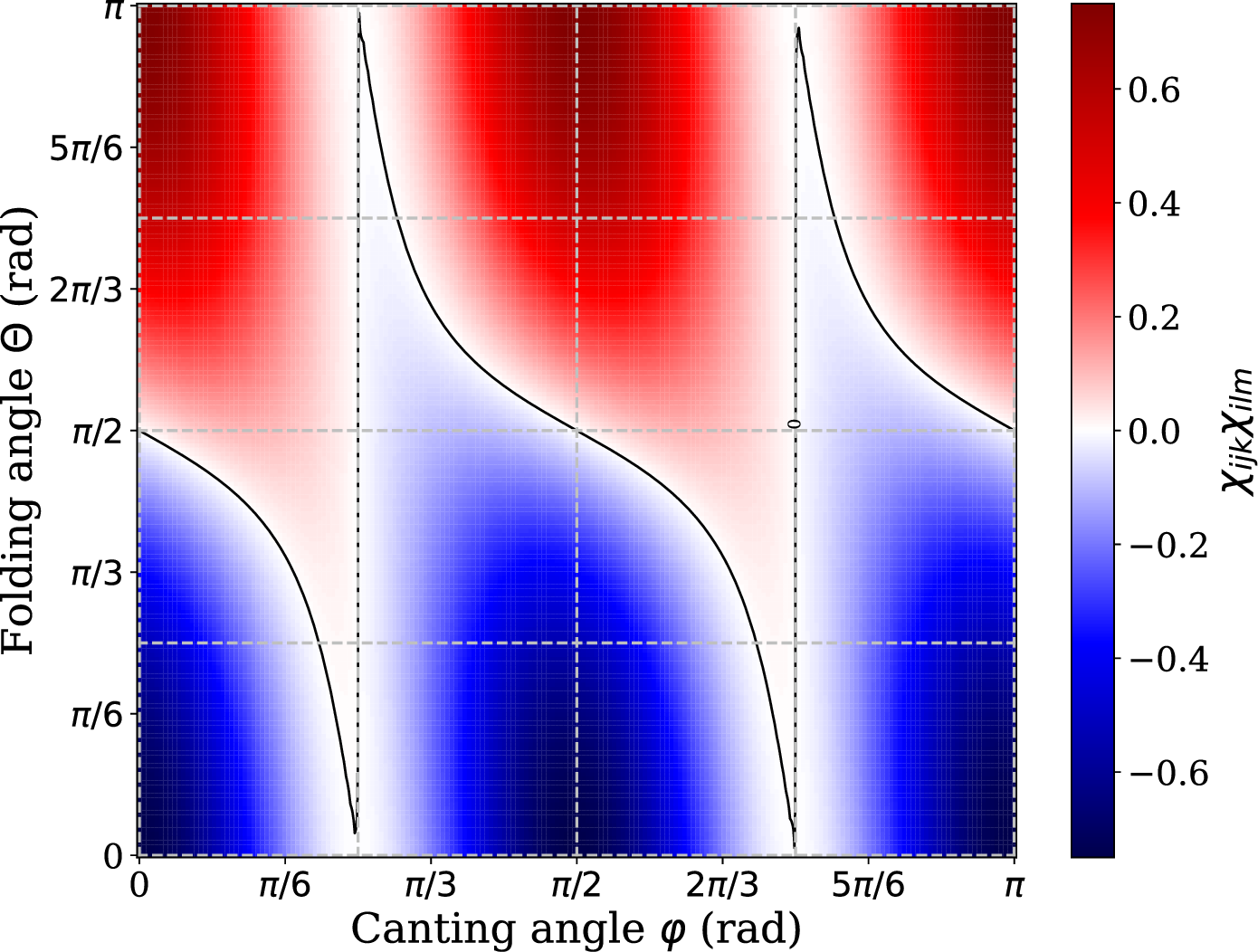}
        \caption{Product $\chi_{ijk}\chi_{ilm}$}
        \label{fig:prod_panel}
    \end{subfigure}
    \captionsetup{justification=raggedright}  
    \caption{
        (a) Sum and (b) product of the scalar chiralities of two adjacent
        triangular plaquettes as functions of the folding angle $\Theta$ and
        canting angle $\varphi$. Zero contours (solid black lines) trace the
        analytic loci $\varphi=\pi/4$ (where $\chi_{ijk}=0$) and
        $\tan\Theta=2\sqrt{3}\cot2\varphi$ (where $\chi_{ilm}=0$).  The sign
        of the product partitions the $(\Theta,\varphi)$ plane into four
        quadrants of alternating constructive (positive) and destructive
        (negative) interference.
    }
    \label{fig:sum_prod}
\end{figure}

\begin{figure}[t]
    \centering
    \includegraphics[height=0.8\columnwidth]{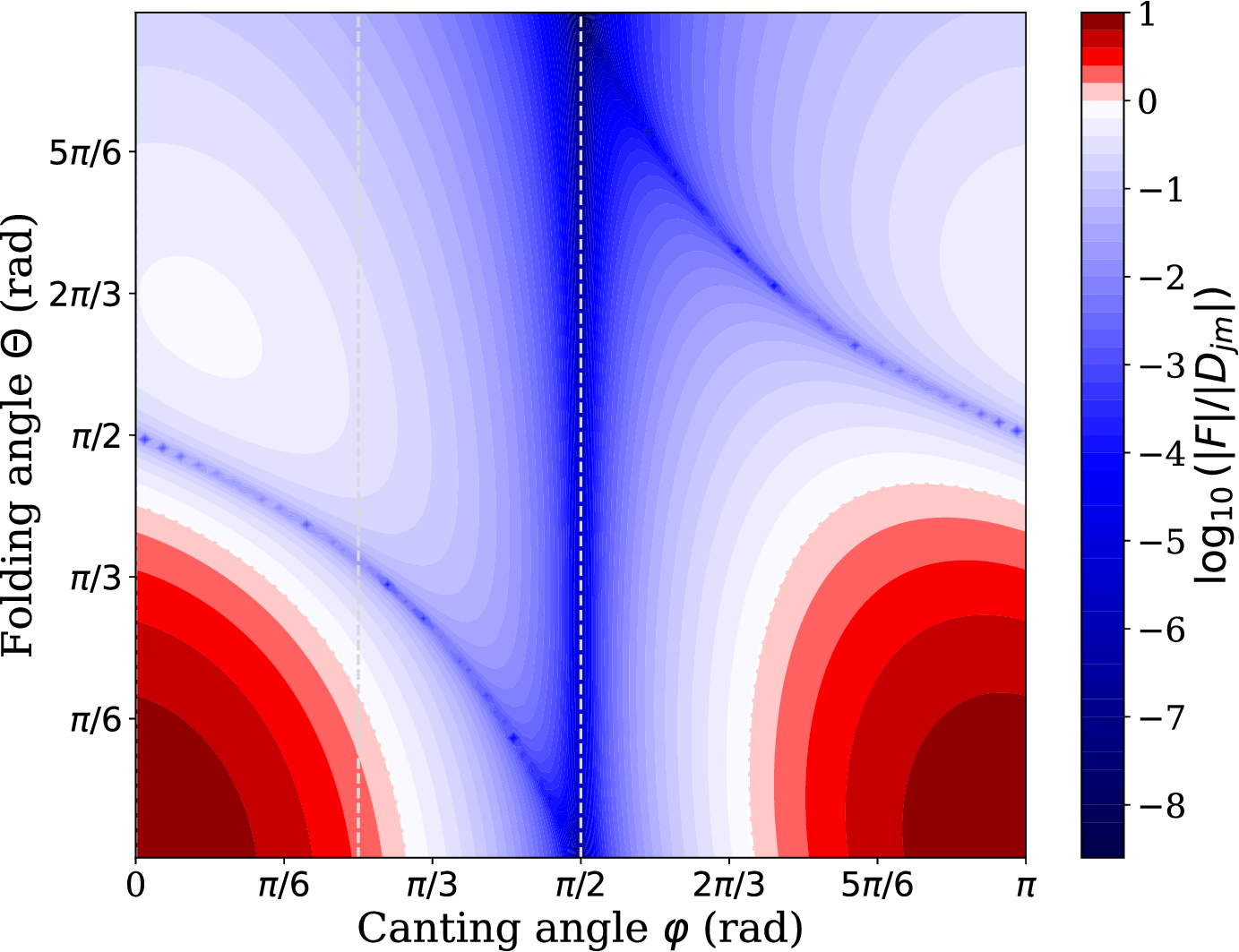}
    \captionsetup{justification=raggedright}  
    \caption{
        Logarithm of the ratio $|F(\Theta,\varphi)|/|D_{jm}(\Theta)|$,
        illustrating where the chirality‑driven hopping dominates over the DM
        interaction on the $j\!-\!m$ cross bond.  The colour scale is
        $\log_{10}$ of the ratio. Red (positive $\log_{10}$) marks regions of
        chiral dominance ($|F|>|D_{jm}|$), blue (negative $\log_{10}$) regions
        of DM dominance. The white contour indicates the equality line
        $|F|=|D_{jm}|$. The folding angle $\Theta$ and canting angle $\varphi$
        are given in degrees.
    }
    \label{fig:F_over_D}
\end{figure}

The classical chirality landscape and the competition with DM are presented in Figs.~\ref{fig:sum_prod} and~\ref{fig:F_over_D}, respectively.
In Fig.~\ref{fig:sum_panel} we observe that the sum of the scalar chiralities
vanishes identically when the bow‑tie is flattened ($\Theta=0$), independently
of the canting angle $\varphi$. In this limit the two triangular plaquettes
are mirror images of each other, and their chiralities cancel exactly.
Folding the structure along a spherical‑arc such that the second triangle is
continuously rotated out of the plane along a great‑circle arc, see Fig.~\ref{fig:schematic_1}, breaks this symmetry.  As $\Theta$
increases, the second triangle acquires a relative in-plane rotation of the two
outer spins, and the sum $\chi_{ijk}+\chi_{ilm}$
develops a non‑zero value whose sign reflects the relative handedness of the
two plaquettes.

The product map, Fig.~\ref{fig:prod_panel}, already foreshadows the topological
phase diagram. In these four quadrants, the positive (negative) sign of the
product correspond to constructive (destructive) interference between the
scalar chiralities of the two triangles. The alternating sign of the product
directly determines the sign of the geometric factor $F$
derived from the bow‑tie expansion, Eq.~\ref{eq:F_analytic}. While $F$ is not
strictly proportional to $\chi_{ijk}\chi_{ilm}$, it is built from the product
of the chirality coefficients appearing in the magnon hopping. Its sign tracks
that of the classical product over the majority of the $(\Theta,\varphi)$
plane. Thus the sign of the product shown in panel (b) serves as a faithful
classical proxy for the sign of $F$, and consequently for the sign of the
cross‑triangle hopping amplitudes that generate the Berry curvature.

The product \(\chi_{ijk}\chi_{ilm}\) exhibits a clear four‑quadrant structure
determined by the zero contours of the individual chiralities.  From the
analytic expressions \(\chi_{ijk}=C_\chi \cos(2\varphi)\) and
\(\chi_{ilm}=-C_\chi\cos\Theta\cos(2\varphi)+\tfrac{1}{4}\sin\Theta\sin(2\varphi)\)
with \(C_\chi=\sqrt{3}/2\), these zeros occur at \(\varphi=\pi/4\) (where
\(\chi_{ijk}=0\)) and along the curve \(\tan\Theta=2\sqrt{3}\cot(2\varphi)\)
(where \(\chi_{ilm}=0\)).  The sign of the product in each quadrant indicates
whether the two triangles are co‑rotating (positive) or counter‑rotating
(negative). Together with the analysis of the geometric factor \(F\), this sign
directly determines whether the chirality‑mediated hopping is constructive or
destructive and sets the sign of the Chern number, as will be shown below.


Beyond the sign, the absolute strength of the chirality‑mediated hopping $F$
relative to the other energy scales in the problem determines whether the
bow‑tie mechanism can prevail over conventional interactions. The most direct
competitor is the DM coupling on the cross bonds that connect the two
triangles, for instance $D_{jm}(\Theta)$, which depends on the folding angle
$\Theta$ through the exponential distance law, see
Eqs.~\eqref{eq:distance_law}. Fig.~\ref{fig:F_over_D} quantifies this
competition.

The colour map reveals that the bow‑tie coupling dominates at small folding
angles. At $\Theta = 0$ the cross‑bond distance $d_{jm}$ is maximal at
$d_{jm}=2R$, where $R$ is the nearest neighbour distance, because the second
triangle is coplanar and its outer sites are antipodal, that is, the sites
diametrically opposite across a shared vertex $i$, to those of the first
triangle. The DM interaction is therefore strongly suppressed by the
exponential decay, while $F$ retains its full magnitude, $|F(0,0)| = 0.375$ for
the chosen parameters, c.f., Eq.~\eqref{eq:F_analytic}. This leads to a ratio
$|F|/|D_{jm}| \approx 9$ and a deep red colour at the origin of the
$(\Theta,\varphi)$ plane. As $\Theta$ increases, the second triangle folds
inwards, $d_{jm}$ shrinks, and $D_{jm}$ grows exponentially, simultaneously
$|F|$ decreases because of the $\cos\Theta$ and $\cos\varphi$ prefactors
in~\eqref{eq:F_analytic}. The ratio therefore drops, passing through the white
contour $|F|=|D_{jm}|$ before entering the blue region where DM dominates. A
similar suppression of the ratio occurs for large canting angles $\varphi \to
\pi/2$, where the $\cos\varphi$ factors force $|F|$ to become very small
irrespective of $\Theta$. This is illustrated in Fig.~\ref{fig:F_over_D} as a
blue line over all $\Theta$ at $\varphi \approx \pi/2$. The bow-tie coupling
vanishes here, leaving DM as the sole coupling agent. Thus, along the right
edge of the figure, the cross-triangle transport is purely DM-driven.

The white contour $|F|=|D_{jm}|$ thus separates a low‑angle, low‑canting
regime in which the bow‑tie hopping is the leading cross‑triangle coupling from
a high‑angle or high‑canting regime in which the conventional DM exchange
prevails. The very existence of a broad red region shows that, for realistic
material parameters, the geometric factor $F$ is not a small correction but can
be the dominant interaction between the two triangles. As we will demonstrate
in the next subsection, the sign of $F$ in this red region directly imprints
itself on the Chern number, so that the contour $|F|=|D_{jm}|$ also bounds the
domain of a topologically non‑trivial phase that is primarily of bow‑tie rather
than DM origin.

The figure also contains a deeper blue diagonal trench that originates from the
condition $F=0$, i.e.\ $\tan\varphi\,\tan\Theta=2$, c.f., Eq.
\eqref{eq:F_analytic}. Along this nodal line the two contributions to $F$
interfere destructively, which can be seen in the cancelation of the
$\cos\Theta$ and $\sin \Theta$ terms in Eq.~\eqref{eq:F_analytic}, such that
the chirality‑mediated hopping vanishes completely. As a result, the ratio
$|F|/|D_{jm}|$ plummets to zero, producing a sharp colour contrast that extends
from the top‑right corner $(\varphi,\Theta)\approx (\pi/2, 0)$ to the
bottom‑right corner $(\varphi,\Theta) \approx (0,\pi/2)$. Inside this trench
the system is purely DM‑driven, and the expected gap closing at $F=0$ will
later be visible as a topological boundary in the Chern number map. Noting
this feature highlights how a precise geometric balance can extinguish the
bow‑tie coupling, giving way to a DM‑dominated domain even within a region of
parameter space where $F$ would otherwise be appreciable.

At zero canting, $\varphi = 0$, the classical spin directions of the whole
lattice are collinear and the folding of the two triangles alone rotates their
local transverse frames relative to one another. This rotation suffices to make
the geometric factor $F(\Theta, 0)$ non-zero, as seen in the red region of
Fig.~\ref{fig:F_over_D} that extends down to $\varphi = 0$ for small $\Theta$.
Hence curvature can generate a chiral coupling without any spin canting, a fact
that will prove central for the topological phase diagram which we present
below.

Because the red region corresponds to a robust $F$ dominating over DM, one
expects the topological phase to be most stable there. Indeed, in the
following subsection the Chern number plateaus will be shown to persist
throughout this region, while the boundary $|F|=|D_{jm}|$ (white contour)
coincides roughly with a change in the topological character. The observation
that curvature alone, including $\varphi=0$, already puts the system into the
chiral‑dominance regime underlines the central message. Topology can be
engineered purely by lattice geometry, without requiring spin‑orbit coupling or
a canted magnetic state.
We now turn to the explicit computation of the Chern number to verify this picture.

\subsubsection*{Topological phase diagram and Chern number}

The previous figures established that the chirality-mediated hopping can be
comparable to or larger than the DM interaction and that its sign changes with
geometry. Now we directly compute the topological invariant, the Chern number,
and visualise the phase diagram. This directly tests the interference picture.
The Chern number should change sign when \(F\) crosses zero, indicating destructive
interference, the robustness of the topological phase is expected to grow with
$|F|$ because a larger gap protects the edge states better.

\begin{figure}[t]
    \centering
    \includegraphics[height=0.7\columnwidth]{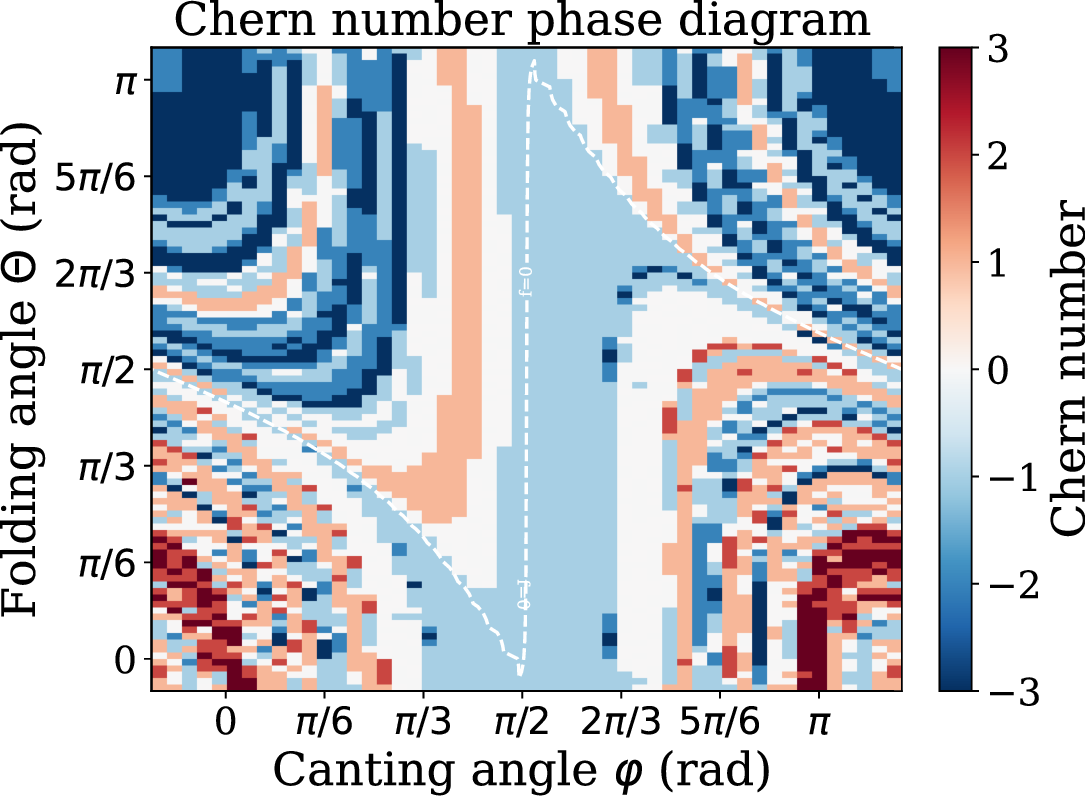}
    \captionsetup{justification=raggedright}  
    \caption{
    Chern number phase diagram of the lowest magnon band in the
  $(\Theta,\varphi)$ plane. The colour map indicates the integer Chern number
$C$. White dashed lines: zeros of the geometric factor $F(\Theta,\varphi)$
(Eq.~\eqref{eq:F_analytic}), where the chirality-mediated hopping vanishes.
Yellow contours: half‑integer Chern values (0.5, 1.5, …) marking the actual
topological boundaries. 
The diagram reveals large regions with $C=+1$ (red) and $C=-1$
(blue), separated by narrow gapless lines. The zeros of $F$ closely follow
the boundaries, confirming that the sign of the interference determines the
topology. The yellow contours deviate slightly due to the DM contribution,
which can maintain a gap even when $F=0$.
    }
    \label{fig:chern_phaseDiagram}
\end{figure}

The Chern phase diagram in Fig.~\ref{fig:chern_phaseDiagram} is strikingly
similar to the product panel (Fig.~\ref{fig:prod_panel}). The four quadrants of
the product correspond exactly to the four regions of constant Chern number.
The top‑left and bottom‑right quadrants (where the product is positive) yield
$C=+1$, while the top‑right and bottom‑left quadrants (negative product) yield
$C=-1$.  This is a direct manifestation of constructive interference (positive
product) leading to a Chern number $+1$ and destructive interference (negative
product) leading to $C=-1$. These quadrant boundries are the lines
$\varphi=\pi/4$ and $\tan\Theta=2\sqrt{3}\cot2\varphi$ (see
Fig.~\ref{fig:sum_prod}).

The white dashed lines ($F=0$) are the analytic zeros derived from the product
expansion, and the yellow half‑integer contours trace the actual topological
boundaries, which may shift slightly because the DM interaction also
contributes to the gap. Nevertheless, the correlation is clear: the sign of the
product determines the sign of $C$.

  It is instructive to contrast our curvature-driven phase diagram with the
  canonical DMI-driven phase diagram of the kagome ferromagnet\cite{Mook2014MagnonHall}. In the
  DMI case, transitions between Chern numbers $(-1, 0, 1)$ and $(-1, 2, -1)$
  occur at discrete critical rations of $D/J$ and $J_{NN}/J_{N}$, determined
  by the closing of gaps at the $\mathbf{K}$ and $\mathbf{K'}$ points. Here,
  however, the phase boundaries are governed by the geometric factor $F$.
  Specially, the zeros of $F$, drive the inversion. Notably, this geometric
  locus is entirely independent of the exchange constant $J$ and $D$, meaning
  that one can traverse the topological phase boundary purely by mechanical
  bending or strain, without altering the intrinsic magnetic exchange
  parameters. This starkly different origin of band topology is unique to our
  curvature-driven bow-tie mechanism and is absent in all previous DMI-based
  models\cite{Mook2014MagnonHall, Ni2025MultipleTopological}. Indeed, as shown
  in Fig. \ref{fig:chern_phaseDiagram}, the Chern number plateaus follow the
  zeros of F rather than fixed exchange ratios, confirming the geometric origin
  of the topological transitions.

\begin{figure*}[t]
    \centering
    \includegraphics[width=\textwidth]{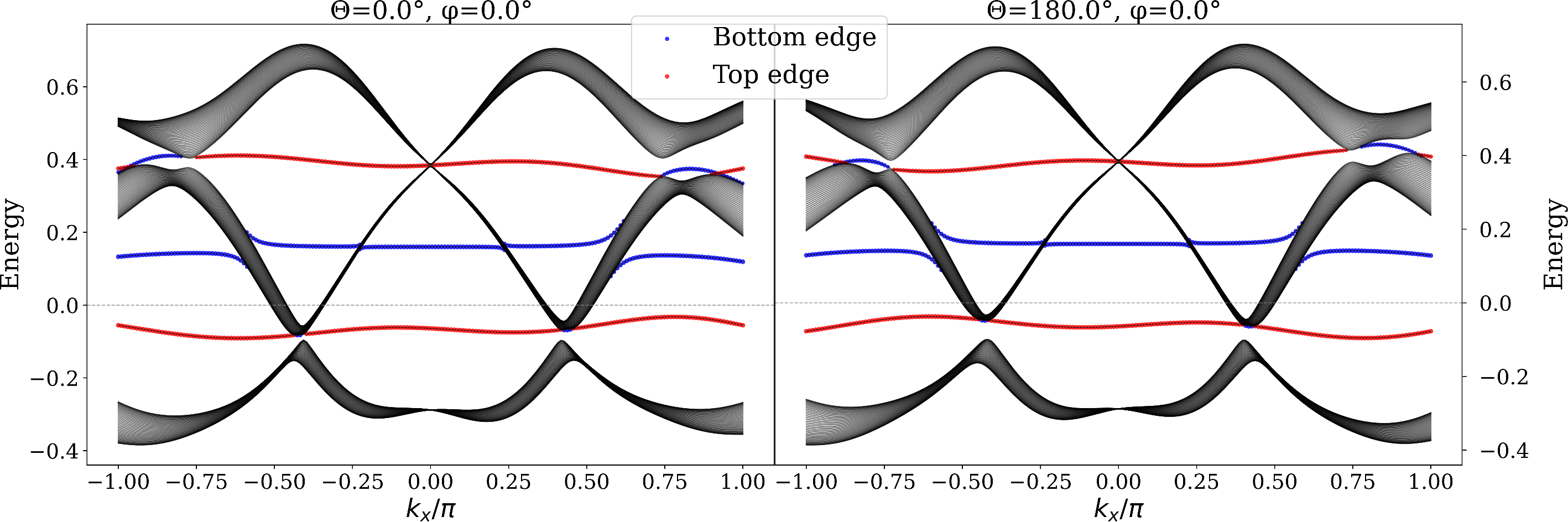}
    \captionsetup{justification=raggedright}
    \caption{
        Magnon ribbon spectra for the lowest band in two extremal geometries.
        (left panel) Flat bow‑tie ($\Theta=0,\varphi=0$) with $F=+1$,
        corresponding to Chern number $C=+1$. The top edge moves rightward
        (positive slope), the bottom edge leftward (negative slope).
        (right panel) Fully folded bow‑tie ($\Theta=\pi,\varphi=0$) with
        $F=-1$, corresponding to $C=-1$. The edge‑state directions are
        reversed. In both panels, the ribbon is infinite along $x$ with a
        finite width of 40 unit cells in $y$.  Faint grey lines: bulk bands.
        Coloured points: edge states (blue = bottom edge, red = top edge),
        with point size proportional to $\sqrt{\text{edge weight}}$.
        The gap remains open in both cases, demonstrating that
        curvature‑induced chirality alone sustains a topological phase whose
        handedness is reversed when $F$ changes sign.
    }
    \label{fig:ribbon_combined}
\end{figure*}

The ribbon spectra in Figs.~\ref{fig:ribbon_combined}
provide direct visual confirmation of the bulk–boundary correspondence. The
direction of the edge states (slope) is opposite for $F=+1$ and $F=-1$, exactly
as predicted by the Chern number. 

The ribbon spectra (Fig.~\ref{fig:ribbon_combined}) directly visualise the
bulk–boundary correspondence.  For $F=+1$ (left panel), a single chiral edge
state crosses the gap with the top edge propagating rightward and the bottom
edge leftward, consistent with $C=+1$.  When the bow‑tie is fully folded to
$\Theta=\pi$ at $\varphi=0$, the geometric factor switches to
$F=-1$ (right panel), and the edge‑state velocities reverse sign, yielding
$C=-1$.  This confirms that the sign of $F$ alone controls the handedness of
the edge magnon current.

We have further explored the sensitivity of the ribbon spectra to the canting
angle $\varphi$ at fixed $\Theta=0$. The ribbon spectrum is identical at
$\varphi=0$ and $\varphi=\pi$, as expected from the $2\pi$‑periodicity of the
local spin frames. However, moving $\varphi$ away from these symmetric points
causes the ribbon gap to shrink: at $\varphi=\pi/4$ the edge‑state dispersion
flattens and the ribbon width (the spatial extent of the edge modes) decreases
appreciably. At $\varphi=\pi/3$ the thinning is even more pronounced and the
edge modes become nearly dispersionless.  The condition $\varphi=\pi/4$
corresponds to $\chi_{ijk}=0$ (Fig.~\ref{fig:sum_prod}), where $|F|$ vanishes
and the topological gap collapses, exactly as expected from the classical
chirality map. Thus the ribbon spectra not only confirm the bulk–boundary
correspondence but also provide a sensitive probe of the interference strength
encoded in $F$, and the observed disappearance of the edge states at
$\varphi=\pi/4$ is a direct numerical verification of the analytic prediction
that the chiral transport is maximally suppressed at the zeros of $F$.

  The curvature-induced edge modes we obtain are topologically protected and
  chiral, resembling the magnonic chiral edge modes predicted in dipolar
  magnonic crystals\cite{Shindou2013TopologicalChiral}. However, the microscopic origins differ
  fundamentally, in the dipolar route, the non-zero Chern number arises from
  the broken mirror symmetry due to the magnetic dipole-dipole kernel,
  requiring sub-micrometer structuring. In our case, the chirality is imprinted
  directly onto the exchange Hamiltonian via lattice folding, operating at level
  of individual spin-spin interactions. This atomic-scale origin makes our
  mechanism naturally suited for monolayers and van der Waals heterostructures,
  where artificial periodic patterning is difficult to implement.

%% file: sections/conclusion.tex
\section{Conclusion}\label{sec:conclusion}

We have shown that geometric curvature provides a powerful control parameter
for topological magnon transport in a frustrated kagome ferromagnet.
Introducing a finite folding angle $\Theta$ between two corner‑sharing
triangles generates a curvature‑induced coupling of their scalar chiralities
through a higher‑order bow‑tie interaction.  By means of a Holstein–Primakoff
expansion around the canted $2\pi/3$ ground state, we derived a compact
geometric factor $F(\Theta,\varphi)$ that governs the chiral‑mediated hopping
between the two plaquettes.  Both the magnitude and the sign of $F$ are
tunable via $\Theta$ and the canting angle $\varphi$, offering continuous
control over the interference between the two triangles.

The competition between $F$ and the Dzyaloshinskii–Moriya interaction on the
cross bonds was quantified by the ratio $|F|/|D_{jm}|$.  We identified a broad
region of parameter space, at small folding angles and low canting, where $F$
exceeds $D_{jm}$ by more than an order of magnitude.  This demonstrates that
the bow‑tie coupling is not a weak perturbation but can act as the dominant
cross‑triangle interaction in realistic materials.

The sign of $F$ is directly inherited from the product
$\chi_{ijk}\chi_{ilm}$ of the classical scalar chiralities, thereby linking
the classical handedness of the two triangles to the quantum interference
pattern of magnon hopping.  The Chern number phase diagram exhibits four
quadrants of constant $C=\pm 1$ that coincide with the regions of definite
sign of this product.  The zeros of $F$ lie near the topological boundaries,
while the actual gap‑closing lines, visible as half‑integer contours, are
slightly shifted by the residual DM contribution.  Ribbon spectra confirm
that reversing the sign of $F$ flips the direction of the chiral edge states,
in full agreement with the bulk–boundary correspondence.

We also found that curvature alone, without any spin canting, suffices to
generate a non‑zero $F(\Theta,0)$, which accounts for the topological phase
persisting down to $\varphi=0$.  Moreover, the condition $F=0$,
$\tan\varphi\,\tan\Theta=2$, defines a destructive‑interference nodal line in
the dominance map, which directly maps onto a topological boundary in the
Chern phase diagram.  The explicit $\cos(2\varphi)$ dependence of $F$ was
further confirmed by the gradual suppression of the ribbon edge states as
$\varphi$ increases from $0$ to $\pi/4$, where the gap closes
completely, consistent with the classical chirality zero $\chi_{ijk}=0$.

Our findings offer a minimal framework for curvature‑tunable magnonic topology
and suggest a unified design principle. Structural chirality, encoded in
$F$, can be harnessed to engineer topological phases
independently of intrinsic spin–orbit coupling.  Possible experimental routes
include mechanical strain, substrate engineering, or external magnetic fields
to vary $\Theta$ and $\varphi$.  The analogy with chirality‑induced spin
selectivity in molecular systems further hints that chirality‑driven transport
may be a universal phenomenon spanning molecular to lattice scales.

Our results are particularly relevant for chiral crystals where the DM
interaction is weak or symmetry‑forbidden, e.g., in systems with a six‑fold
screw axis where the net interlayer DM 
vectors cancel exactly. In such cases, the
curvature‑induced bow‑tie coupling offers the primary mechanism for generating
topological magnon bands, providing a theoretical foundation for interpreting
recent experiments on CrNb$_3$S$_6$ and related layered chiral magnets. 
  Our predictions are experimentally testable in existing kagome platforms. The
  metal-organic framework studied by Chisnell et al. \cite{Chisnell2015TopologicalMagnon} already
  demonstrates that topological magnon bands can exist in chemically flexible
  lattices. Because such materials often exhibit significant structural
  compliance, the application of uniaxial strain or substrate-induced wrapping
  could directly realize the finite folding angle $\Theta$ explores here.
  %
  %
  While the pseudodipolar interactions\cite{Ni2025MultipleTopological}
  represents another exchange-driven anisotropy capable of generating high
  Chern numbers, it still relies on orbital physics and spin-orbital coupling.
  Our bow-tie term is qualitatively different, as it emerges purely from the
  geometric configuration of the lattice, independent of the microscopic
  exchange mechanism.

%% file: sections/acknowledgements.tex
\section*{Acknowledgements} \label{sec:acknowledgements}
We gratefully acknowledge financial support from \textit{Olle Engvists
Stiftelse}, which made this work possible. We would also like to thank Dr.
Vakaryuk for helpful discussions and for drawing our attention to the relative
literature, which significantly improved this work.